\documentclass[prd,twocolumn]{revtex4}
\usepackage{graphicx,verbatim,amsmath}

\def\beq{\begin{equation}}
\def\eeq{\end{equation}}
\def\bea{\begin{eqnarray}}
\def\eea{\end{eqnarray}}

\def\d{\partial}
\def\sgn{\epsilon}
\def\detgam{\sqrt{|\gamma|}}

\newcommand{\Lie}{\pounds}

\begin{document}

\title{Self-Renormalization of the Classical Quasilocal Energy}

\author{Andrew P. Lundgren}
\affiliation{Department of Physics, Cornell University, Ithaca, New York 14853}
\email{apl27@cornell.edu}

\author{Bjoern S. Schmekel}
\affiliation{Theoretical Astrophysics Center, University of California, Berkeley, California 94720}
\email{schmekel@berkeley.edu}

\author{James W. York, Jr.}
\affiliation{Department of Physics, Cornell University, Ithaca, New York 14853{}}
\email{york@astro.cornell.edu}

\begin{abstract}
Pointlike objects cause many of the divergences that afflict physical theories.  For instance, the gravitational binding energy of a point particle in Newtonian mechanics is infinite.  In general relativity, the analog of a point particle is a black hole and the notion of binding energy must be replaced by quasilocal energy.  The quasilocal energy (QLE) derived by York, and elaborated by Brown and York, is finite outside the horizon but it was not considered how to evaluate it inside the horizon.  We present a prescription for finding the QLE inside a horizon, and show that it is finite at the singularity for a variety of types of black hole.  The energy is typically concentrated just inside the horizon, not at the central singularity.
\end{abstract}

\maketitle

\section{Introduction}

It is a fundamental fact of general relativity (GR) that there is no such concept as the local energy of the gravitational field.  The local effects of gravity can be removed by transforming to a freely falling frame.  A neutral object at the origin of a freely falling frame will not experience any gravitational acceleration.  Tidal forces will remain, but they only act on particles that are separated by some distance.  Since gravity has no local effect, there exists no local energy.

Various definitions of local energy densities can be made by making reference to special coordinate systems or background metrics.  Heuristically, if we have an observer that we consider static we could use its acceleration as a measure of the local gravity.  More sophisticated approaches yield a variety of quantities that are useful for certain applications.  The structure of GR is such that local quantities representing an energy do not exist, therefore any attempt to define them must use use concepts that are not a natural part of the theory, i.e., a special coordinate system.  Studying the asymptotic behavior of the metric, as done in post-Newtonian approximations, or the behavior of the metric at spatial or null infinity \cite{adm, misner, landau, bondi} leads to more useful and natural formulas for the energy.  From these ideas, we are led to the idea of finding the energy inside a given finite region rather than the energy at a point.

Quasilocal energy (QLE) is the energy inside a two-dimensional surface.  The surface could be a sphere enclosing a star or a black hole, a small box enclosing some matter undergoing cosmological expansion, or a complicated, even disconnected, surface in the spacetime.  In this paper, we follow the method of Brown and York \cite{brown1993,brown1998} which derives an energy from a Hamilton-Jacobi argument involving the canonical action.  This QLE has many useful properties.  For example, it agrees with the Newtonian limit for a spherical star, is applicable in thermodynamic problems \cite{york1986,bbwy}, and the asymptotic limit at Euclidean infinity is the ADM expression for energy.  Furthermore, it can be directly obtained from the Hamiltonian of the same action principle (footnote 14 of \cite{bbwy}) without the need for any other geometric structures.  There are many formulas for other quasilocal energies \cite{szabados2004} (and many references given therein), derived using different methods and often having different properties.

We can define the quasilocal energy of the electric field in classical physics for comparison.  The electric field of a point charge falls off with $1 / r^2$, and the energy density equals the field strength squared.  We have for the energy inside a surface of radius $R$
\beq
E(R) \propto \int_0^R (\frac{1}{r^2})^2 r^2 dr = \frac{1}{0} - \frac{1}{R}
\eeq
which makes an infinite contribution because the charge is pointlike.  This problem remains in quantum electrodynamics and requires renormalization, where another infinite quantity is subtracted to leave a finite remainder.

In GR, the situation is somewhat different because objects of a given mass cannot have an arbitrarily small size.  Once they become too small, they collapse to form a black hole and an event horizon forms.  The resulting object is effectively the size of the event horizon, and outside observers are shielded from the infinities at the center.  The QLE for a Schwarzschild black hole of mass $M$ has the large distance limit
\beq
E(R) = R \left(1 - \sqrt{1- \frac{2M}{R}}\right) \approx M + \frac{M^2}{2 R} .
\eeq

The QLE becomes undefined at $R \leq 2M$, the radius of the event horizon.  The large distance limit suggests that the energy will diverge at the center, although we are protected from seeing this behavior by the event horizon.  However, this is still something of a problem because an observer can fall in through the horizon in finite proper time and survive to see the interior of the black hole.  It may be useful to have a definition of energy for observers inside the event horizon.

A major issue that arises when defining QLE is which observers to use.  Different observers, for instance ones that are accelerated rather than free-falling, will measure a boosted energy.  In this paper, we choose the observers that are stationary with respect to the boundary, i.e., their four-velocity is perpendicular to the normal.  These are the observers that will not stretch or squash the boundary, which would affect the energy.  We will show that this prescription does not depend on the time slicing, and that the QLE so defined has useful properties.  By comparing the QLE with the square root of the area of the boundary (which we refer to as the radius, although they are not equivalent), we can determine whether the radial coordinate is timelike or spacelike.  This is simply the idea that if there is too much energy inside the corresponding Schwarzschild radius, a horizon must form.

The derivation in \cite{brown1993} involved the boundary term of the Hilbert action.  We write this action for a general spherically-symmetric and static metric and show how the QLE formula can be modified so that the boundary term is treated correctly in either case.  Surprisingly, when not coupled to other fields, the energy of the singularity at the center of a black hole is \emph{zero}.  The energy climbs toward a maximum value at a radius inside the horizon, and at the horizon has an infinite downward slope.  The charged black hole has a finite but \emph{negative} energy at the singularity.

\section{The Brown-York Quasilocal Energy}

We now review the quasilocal energy defined in \cite{brown1993}.  The basic idea is the Hamilton-Jacobi method in classical mechanics of expressing the energy as a variation of the action with respect to the endpoints.  The generalization to curved spacetimes results in the following definition
\beq
E = \frac{1}{\kappa} \int_B d^2 x \sqrt{\sigma} \left ( k - k_0 \right )
\eeq
where $\sigma_{\mu \nu}$ is the induced metric on the boundary
\beq
\sigma_{\mu \nu} = g_{\mu \nu} + u_{\mu} u_{\nu} - n_{\mu} n_{\nu}
\eeq
and $\sigma$ is its determinant.  In the last equation $u^{\mu}$ is a future pointing timelike unit normal for the spacelike hypersurface $\Sigma$ whereas $n^{\mu}$ is an outward pointing spacelike normal to the boundary $^3 B$ which is also normal to $B$ if $u \cdot n = 0$ which is assumed in this definition.  The constant $\kappa = 8 \pi G$ is just a constant of proportionality, and in natural units is just $8 \pi$.

The $k$ in the above equation is the trace of the extrinsic curvature of the two-boundary's embedding into the spacelike hypersurface $\Sigma$.  The $k_0$ term is the energy of the vacuum, which must be subtracted to obtain the physical energy.  In our case, we take flat space as the vacuum so that $k_0$ is the trace of the extrinsic curvature for the same two-boundary embedded in flat space.   It is natural to add such a term because otherwise the intrinsic geometry of the surface would contribute to the energy, even with no gravitational energy present.

$E(R)$ has been computed already in \cite{brown1993} for 4 dimensional spherically symmetric objects outside the event horizon.  The metric of the spacelike slice can be written as
\beq
ds^2 = f(r)^{-2} dr^2 + r^2 ( d\Omega^2 )
\label{4Dmetric}
\eeq
and the unit normal to constant $r$ surfaces is
\beq
n^{\mu} = (0, f(r), 0, 0) .
\eeq

The extrinsic curvature of the two-boundary is
\beq
k_{\mu \nu} = - \sigma^{\alpha}_{\mu} \nabla_{\alpha} n_{\nu}~;~~k = - \sigma^\alpha_\mu \nabla_\alpha n^\mu
\eeq
where the above covariant derivative is taken in the spacelike slice, and $\sigma$ is serving as a projection operator; it can be found from
\bea
\sigma_{\mu \nu} &=& h_{\mu \nu} - n_{\mu} n_{\nu} \\
\sigma_{\mu}^{\nu} &=& h^{\nu \rho} \sigma_{\rho \mu}  = \delta_\mu^\nu - n_\mu n^\nu
\eea
where $h_{\mu\nu}$ is the metric of the spacelike slice \eqref{4Dmetric}.  The only connection coefficients that we need are
\beq
\Gamma^{\theta}_{r \theta} = \Gamma^{\phi}_{r \phi} = \frac{1}{r} .
\eeq
and we obtain
\beq
k = - 2 \frac{f(r)}{r} .
\eeq
The reference term $k_0$ comes from setting $f(r) = 1$ so that we are embedding the sphere in flat space.  For a spherical star with energy density $\rho(r)$, we have \cite{MTW}
\bea
f(r) = \sqrt{1 - 2 m(r) / r} \\
m(r) = 4 \pi \int_0^r r'^2 dr' \rho(r')
\eea
or for a black hole we simply have $m(r) = M$ and $f(r) = 1 - 2 M / r$.  With
\beq
\sqrt{\sigma} = r^2 \sin \theta ~,
\eeq
the QLE becomes
\beq
E(R) = R \left[ 1 -  \left( 1 - \frac{2 m(R)}{R} \right)^{1/2} \right]
\eeq
as long as $r$ is a spacelike coordinate.

\begin{figure}
\tiny
\includegraphics[width=3.2in]{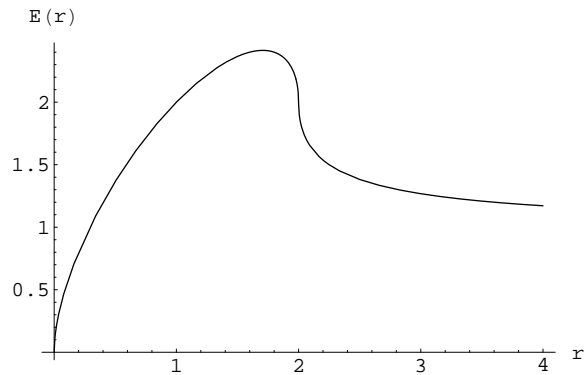}                                          
\caption{QLE computed inside and outside the event horizon for a Schwarzschild black hole.  Both axes are in units of the mass M, and the horizon is at 2M.}
\label{QLEfig}
\end{figure}

Specializing to the Schwarzschild case where $m(r) = M$, there is a horizon at $r = 2 M$.  Inside the horizon the $r$ coordinate becomes timelike as revealed by inspecting \eqref{4Dmetric}.  There is a choice of which sign the unit normal should have inside the horizon; we choose
\beq
n^{\mu} = (0, -(2 M / r -1)^{1/2}, 0, 0)
\eeq
which we will justify later by examining the boundary terms.  The QLE becomes
\beq
E(R) = R \left[ 1 +  \left( \frac{2M}{R} -1 \right)^{1/2} \right]
\eeq
inside the horizon.  The QLE of the entire Schwarzschild metric is plotted in Fig.~\eqref{QLEfig} and shows three striking features.

First, the quasilocal energy at the singularity is zero.  In Newtonian gravity, the energy of the gravitational field would diverge at the center for a point particle.  So the nonlinearity of general relativity has removed this infinity, and gives us a picture where the singularity is not the most important feature of the black hole.  We expected that the mass seen at infinity would reflect the mass of the singularity, but this seems not to be the case.  The black hole looks like an extended object when we consider the second feature, that the QLE attains its maximum inside the horizon at a radius of $1+\frac{\sqrt{2}}{2} M$.  Most of the energy of the black hole seems to be ``stored'' just inside the horizon.

The third striking fact is that the derivative of the QLE matches across the horizon, but is infinite there.  The energy should be continuous on physical grounds, but the derivative might not be.  In fact, if we had chosen the opposite sign of the normal, there would be a cusp in the QLE at the horizon.  We will show that this is the correct choice in the next section.

We note in passing that in three-dimensional spacetime the QLE is constant everywhere, so these features would be missing.  In $2+1$ spacetime, there is no horizon and the metric only possesses a conical defect at the center.  This is because the Schwarzschild metric in $n$ dimensions is
\beq
ds^2 = - \left( 1 - \frac{2 m}{r^{n-3}} \right) dt^2 + \left( 1 - \frac{2 m}{r^{n-3}} \right)^{-1} dr^2 + r^2 d\Omega^2
\eeq
In three dimensions with $d\Omega^2 = d \theta^2$ and $\sqrt{\sigma} = r$ the QLE is constant everywhere.

\section{Relationship between action and QLE}

The above calculation is not convincing because it is not clear that we have chosen the correct unit normal.  A different choice would change the sign and give a much different result.  To fix this ambiguity, we will go back to the derivation of QLE from the action.  The extrinsic curvature formula for the QLE arises from the boundary term in the action.  From the behavior of this term inside and outside the horizon, we can determine how to modify the QLE.

We begin by considering a region of spacetime $M$.  The spacetime is foliated by spacelike hypersurfaces $\Sigma$, which can play the role of moments of time.  The region of spacetime will be bounded by hypersurfaces that we will simply refer to as $t_1$ and $t_2$.  Each $\Sigma$ has a spatial boundary $B$, which in this paper will always have the intrinsic geometry of a sphere.  The time history of the boundaries $B$ will be called the three-boundary $^3B$.

The three-boundary $^3B$ is the important boundary for the derivation of the quasilocal energy.  In Brown and York's derivation, this boundary must be perpendicular to the time slices $\Sigma$.  Besides simplifying the calculation, this is also an important physical point: the observers whose velocities are normal to the time slice are the observers that will measure the QLE.  These observers should be at rest with respect to the boundary, and therefore the time slices should be perpendicular to the boundary.  If they are not, then the observers will be boosted, and it is to be expected that they will measure a different value for the QLE.  In this paper, we will relax the condition slightly but measure the same energy.

To make clear how the boundary term gives rise to the QLE, we will restrict our attention to only a simple class of metrics.  Spherical symmetry is imposed so that we can easily embed a sphere with a given surface area in the four-dimensional metric.  The metrics we study are static, so that the only nonzero derivatives are radial.  While we would like to generalize this derivation at a later date, this restricted version gives several interesting results and has the benefit of being easily understood.

A general spherically symmetric and static metric can be put in the form
\beq
ds^2 = - \sgn N(r)^2 dt^2 + \sgn f(r)^{-2} dr^2 + r^2 ( d\theta^2 + \sin(\theta)^2 d\phi^2 )
\eeq
where $\sgn$ is $1$ outside the horizon and $-1$ inside.  $N$ and $f$ will be chosen to be positive.

To investigate the properties of the action, we will impose boundary conditions at some fixed $r$.  The boundary term that we add is the one suitable to fix the induced metric on the three-boundary $^3B$.  The induced metric in this simple case is
\beq
\gamma_{ij} dx^i dx^j = - \sgn N(r)^2 dt^2 + r^2 ( d\theta^2 + \sin(\theta)^2 d\phi^2 ) .
\eeq

The action, with a boundary term added to fix the metric on the boundary, is
\beq
S = \frac{1}{2 \kappa} \int_M d^4 x \sqrt{-g} R - \frac{\sgn}{\kappa} \int_{\d M} d^3 x \detgam \Theta
\eeq
where the extrinsic curvature is
\beq
\Theta_{ij} = - \frac{1}{2} \Lie_n \gamma_{ij}
\eeq
and the trace $\Theta$ simplifies to
\beq
\Theta = \gamma^{ij} \Theta_{ij} = - \frac{f}{\detgam} \d_r (\detgam) ~.
\eeq

The bulk term for the action using this ansatz for the metric is
\bea
& & S_{bulk} = \frac{1}{\kappa} \int d^4x \sin{\theta}\\
&\times& \left[ N / f -  \sgn ( N f + 2 r N f' + 2 r f N' + r^2 f' N' + r^2 f N'') \right] \nonumber
\eea
where primes denote $r$ derivatives.  The boundary term is
\beq
S_{BT} = - \frac{\sgn}{\kappa} \int d^3x \sin{\theta} ( -f (N r^2)' ) .
\eeq

The boundary term can be converted into an integral over all 4 dimensions by also fixing the metric at $r = 0$ and integrating the derivative of the boundary term.
\beq
S_{BT} = \int d^3x (BT) = \int d^3x dr (BT)' + \left. \int d^3x (BT) \right|_{r=0} \label{FundThmCalc}.
\eeq

Adding the two terms shows that in this case, the action has a very simple form.
\beq
S = \frac{1}{\kappa} \int d^4x \sin(\theta) \left[ (\frac{1}{f} + \sgn f) N  + (2 \sgn  r f) N' \right] .
\eeq
We have dropped the constant that comes from the second term in \eqref{FundThmCalc}, because it will not affect the final result.  Varying $N$ yields
\beq
(2 \sgn r f)' = (\frac{1}{f} + \sgn f) ~,
\eeq
and substituting back into the action shows that the action is an integral of a total derivative.  Also doing the angular integrations (trivial because of spherical symmetry) gives

\beq
S = \frac{8 \pi}{\kappa} \int (N dt)(\sgn r f) ~.
\eeq

Following \cite{brown1993}, we define the quasilocal energy as minus the second term in parentheses, so
\beq
E(r) = - (\sgn r f) ~.
\eeq

The values for the Schwarzschild metric are
\beq
n(r) = f(r) = \sqrt{\sgn(1 - 2 M /r} .
\eeq
We also need to subtract the energy of flat space, which does not depend on $\sgn$; we are embedding a sphere in flat space where there is no horizon.  The subtraction term just has $k_0 = - \frac{2}{r}$.  This will be the subtraction term used in the entire paper.  The result for the QLE for any metric of the form considered is
\beq
E(r) = r (1 - \sgn f(r) )
\eeq
which reproduces the result in the previous section.

\section{Coordinate Independence}

We can relax the restriction on the form of the metric slightly, and consider what happens when the time coordinate is given an $r$ dependence.  The Brown-York derivation requires that the $t$ and $r$ coordinates be perpendicular at the boundary, a condition which is violated by this transformation.  More general derivations of the QLE have been considered \cite{lau, booth, epp} where this condition is eliminated.  We will not consider this issue in depth, but simply use the transformation to show that our version of the QLE is not coordinate dependent, for coordinate transformations of this type.

If we make the transformation $t = t(T,r)$, then we can write $dt = t_T dT + t_r dr$, where subscripts denote derivatives.  The metric becomes
\bea
ds^2 &=& - \sgn N^2 (t_T dT + t_r dr)^2 + \sgn f^{-2} dr + r^2 d\Omega^2 \\
&=& - \sgn (t_T N)^2 (dT + \frac{t_r}{t_T} dr)^2 + \sgn f^{-2} dr + r^2 d\Omega^2 \nonumber
\eea
written in a $3+1$ form where the foliations are hypersurfaces of constant $r$, which is appropriate for finding the induced metric on the three-boundary.  Two good examples of this form are the Eddington-Finkelstein and Painlev\'{e}-Gullstrand coordinates.  The action becomes
\beq
S = \frac{8 \pi}{\kappa} \int_{\Gamma} (N t_T dT + N t_r dr)(\sgn r f) ~.
\eeq
The integral is taken over a contour $\Gamma$ which holds $r$ fixed, so the final result is that the proper time $N dt$ has been transformed to $N t_T dT$, the proper time written in the new coordinates.  The QLE, $- \sgn r f$, is \emph{not} changed.  We will not relate this quantity to the extrinsic curvature because it is not necessary to our point here.

The definition of the QLE is made of simple components.  We embed a three-boundary in the space in such a way that at some moment of time, it has the intrinsic geometry of a sphere with a specified area.  The geometry of the sphere does not change when moving along the time coordinate of the three-boundary.  The QLE is the change in the action with proper time, both of which are invariants.  The invariance of the result under this particular type of coordinate change is therefore not surprising.

\section{deSitter Space and Black Holes}

The details of the derivation are not changed if a cosmological constant is added.  The term added to the action is
\beq
S_{CC} = \int d^4x \sqrt{-g} ( - 2 \Lambda )
\eeq
which is proportional to $N$.  When the equation of motion obtained by varying $N$ is substituted back into the action, the same result as before is obtained but now $f(r)$ is different.

The metric is
\beq
N(r) = f(r) = \sqrt{\sgn(1 - \frac{\Lambda r^2}{3})} .
\eeq
There is a cosmological horizon at $r = \sqrt{3 \over \Lambda}$.  As expected, the energy continually grows with increasing $r$.  The horizon forms when the energy inside the surface grows larger than
\beq
E(R) = \frac{c^4}{G} R .
\eeq
This is the usual Schwarzschild radius expressed in a different form.

\begin{figure}
\tiny
\includegraphics[width=3.2in]{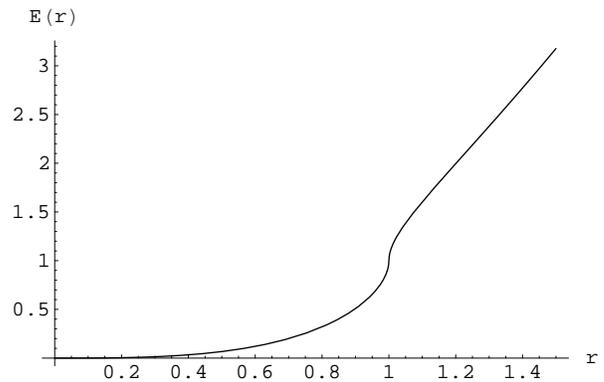}                                          
\caption{QLE of deSitter space (positive cosmological constant).  Both axes are in the same units, with a mass scale proportional to $1 /  \sqrt{\Lambda}$.}
\label{QLEdsfig}
\end{figure}

The deSitter-Schwarzschild solution has both a black hole horizon and a cosmological horizon.  The black hole has a large amount of QLE inside a certain radius, but outside this radius the gravitational binding energy provides enough shielding to bring the energy below $c^4 R / G$.  The inner horizon forms at this radius.  As one gets farther away, the cosmological constant begins to contribute noticeably.  The cosmological horizon forms where the energy has once again climbed above the necessary value.

\begin{figure}
\tiny
\includegraphics[width=3.2in]{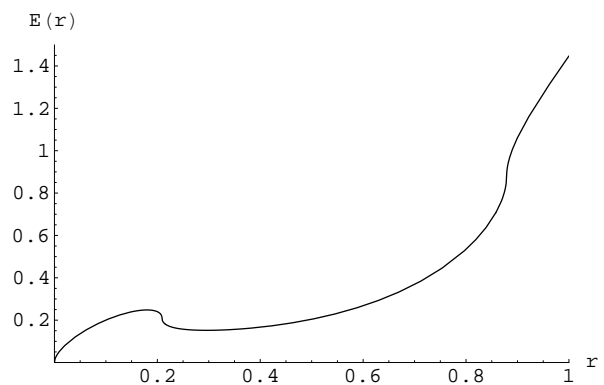}                                          
\caption{QLE of deSitter-Schwarzschild, a black hole in a spacetime with positive cosmological constant.  The units are the same as the plot of the deSitter QLE, and the black hole has an unrealistically large mass so that details can be seen in the plot.}
\label{QLEdsschfig}
\end{figure}

\section{Reissner-Nordstrom}

The Reissner-Nordstrom metric for a charged black hole is interesting because the metric behaves quite differently at the center from the uncharged case.  The form of $N$ and $f$ is now
\beq
N(r) = f(r) = \left(1 - \frac{2M}{r} + \frac{e^2}{r^2} \right)^{1/2} 
\eeq
where $e$ is the charge of the black hole in natural units.  There are now two horizons at $r_\pm = M \pm \sqrt{M^2 - e^2}$.  The outer horizon is the same type as the Schwarzschild horizon.  The inner horizon exchanges the signature of the $t$ and $r$ coordinates again, such that $t$ is a timelike direction.  The consequence is that the singularity is now avoidable.  The inward radial direction is spacelike and not timelike, and so particles are not inexorably drawn into the singularity.

Adding a new field into the theory will not change the definition of the QLE.  The QLE only measures the gravitational energy, and so only the gravitational action is important.  Of course, the addition of a new field changes the metric.  One of the most important characteristics of the gravitational energy is universality.  All mass-energy contributes to gravity, and so the QLE measures the energy of everything inside the surface (including the contributions purely from gravity).

The striking feature of this case is that the energy becomes negative within a certain radius.  The QLE in either region where the time coordinate is timelike is
\beq
E(R) = R \left( 1 - \sqrt{1 - \frac{2M}{R} + \frac{e^2}{R^2}} \right)
\eeq
and so the energy becomes negative for $R < e^2 / 2 M$.  This is always inside the inner horizon.  The energy at the singularity is $E(0) = - |e|$.  The singularity has the electric field of a point charge, and so using just classical electromagnetism, the energy should diverge for small radius.  However, the gravitational binding energy is negative, and while the cancellation is not perfect it seems that the binding energy at least makes the energy at the center finite.

The geodesics of massive neutral particles in the spacetime offer a probe of the effects of negative gravitational energy.  The radial geodesics obey the equation
\beq
\dot r^2 + V(r) = p_0^2 - 1~~; \quad V(r) = (e^2 - 2 M r) / r^2
\eeq
where $p_0$ is the conserved energy per unit mass of the particle, and $\dot r$ is the proper time derivative of $r$.  A particle that starts from $r = \infty$ with zero velocity will not reach the center, but turn around at $r = e^2 / 2 M$.  Particles with higher energies will penetrate farther toward the center, but massive particles of all energies are repelled.  This result is well known \cite{armenti}.  The turnaround radius agrees with the radius where the quasilocal energy becomes negative, so it seems that the two effects are very likely connected.  Negative energy densities are expected to possess repulsive gravitational fields, and negative gravitational energy itself should be no exception.  
\footnote{Negative quasilocal energy has been ruled out by positivity theorems \cite{liuyau} which should apply in this case because the spacetime is static and the time coordinate timelike at the radius we are interested in.  The energy condition which is assumed for the theorem holds true throughout the Reissner-Nordstrom spacetime.  However, the spacelike slice is assumed to be compact within the boundary where the QLE is defined.  We conjecture that this is the condition that breaks down and causes the QLE to become negative.}

\begin{figure}
\tiny
\includegraphics[width=3.2in]{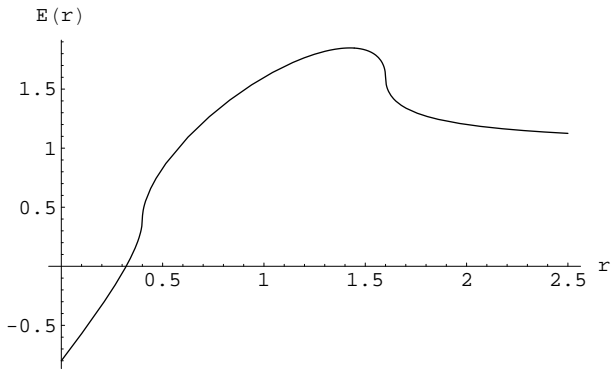}                                          
\caption{QLE of a Reissner-Nordstrom charged black hole.  Both axes are in units of mass of the black hole, and the charge $e^2 = 0.8 M^2$ }
\label{QLEchargefig}
\end{figure}

\section{Conclusion}

We have shown that there is a sensible way to extend the definition of the quasilocal energy to surfaces inside an event horizon.  The Schwarzschild singularity has zero energy, and the energy of the black hole mostly resides in a region just inside the horizon.  The addition of a positive cosmological constant does not change these features but adds a cosmological horizon.  In these cases, when the energy inside a given radius is less than $r$ (in natural units), the space and time coordinates play their usual roles.  When the energy exceeds this quantity, a horizon forms and space and time switch roles.  The derivative of the energy with respect to $r$ at a horizon always seems to be infinite.  These two features make it easy to locate the horizons on a plot of the QLE.

In a Reissner-Nordstrom black hole, the singularity at the center behaves like a point charge, and so there should be a divergence from the positive electric field energy.  However, the contribution from the gravitational binding energy is negative and apparently cancels the divergence, rendering a finite energy at the singularity.  The QLE is negative inside the radius $e^2 / 2 M$, which is always inside the inner horizon of the Reissner-Nordstrom metric.  A massive neutral particle released from rest at infinity will fall to the radius where the energy becomes negative, then reverse direction and be repelled.  This provides the justification for the QLE's negativity, which is also related to the effect that clocks inside this radius run faster than those at asymptotic infinity.

In this paper, we have used a specific preferred set of observers to define the quasilocal energy.  A direction for future research is to remove this restriction as in other work \cite{lau,epp,booth,liuyau} to define a more invariant quantity.  We would like to extend this work to non-spherical boundaries and more general spacetimes.  Two of interest are the spinning black hole metric and the metric for a star collapsing to form a black hole.  There may also be applications to semiclassical and quantum gravity.  Also, the issue of under what conditions the quasilocal energy is negative and what this means physically requires careful attention.

\begin{acknowledgements}
We would like to thank Saul Teukolsky, \'{E}anna Flanagan, Larry Kidder, David Brown, Stephen Lau, Ali Vanderveld, and David Tsang for helpful comments.  JWY thanks the National Science Foundation for support under grant number PHY-0216986.

\end{acknowledgements}

\end{document}